\documentclass[twocolumn,pra]{revtex4}
\usepackage[koi8-r]{inputenc}
\usepackage[english,russian]{babel}
\usepackage{amsmath}
\usepackage{amssymb}
\usepackage{epsfig}

\begin{document}
\title{Узкие световые пучки с линейной поляризацией в керровской среде}

\author{В.\,П.\,Рубан}
\email{ruban@itp.ac.ru}
\affiliation{Институт теоретической физики им.~Л.\,Д.\,Ландау РАН, 142432, 
Черноголовка, Россия}

\date{\today}

\begin{abstract}
В рамках векторного уравнения типа ротор-ротор, описывающего
монохроматическую световую волну в фокусирующей керровской среде,
численно с большой точностью найдена поперечная структура предельно
узких самосфокусированных оптических пучков, имеющих линейную поляризацию. 
Для таких двумерных пространственных солитонов при их ширине всего около
одной длины волны существенны все три компоненты электрического поля. 
Поскольку солитоны являются ``седловыми точками'' некоторого функционала,
использованный здесь численный метод представляет собой релаксационную
процедуру, в которой по устойчивым модам происходит минимизация, а по 
большинству неустойчивых мод --- максимизация функционала. 
Единственная оставшаяся при этом неустойчивая мода общей амплитуды
стабилизируется фиксацией потока энергии через поперечное сечение пучка.

\vspace{4mm}

Key words: optical spatial solitons, Kerr nonlineariry, numerical method
\end{abstract}

\maketitle

{\bf Введение.}
Строго монохроматическая световая волна с частотой $\omega$ в 
пространственно однородной немагнитной оптической среде с керровской 
нелинейностью описывается векторным уравнением \cite{LL8}
\begin{equation}
\mbox{rot}\, \mbox{rot}\, {\bf E}=\frac{\omega^2}{c^2}[\varepsilon{\bf E}
+\alpha |{\bf E}|^2{\bf E}+\beta({\bf E}\cdot{\bf E}){\bf E}^*],
\label{curl_curl}
\end{equation}
которое является одной из наиболее фундаментальных моделей в нелинейной оптике.
Здесь ${\bf E}(x,y,z)$ --- комплексная амплитуда главной гармоники электрического
поля, $\varepsilon$ --- диэлектрическая проницаемость для волн малой амплитуды
на заданной частоте, $\alpha$ и $\beta$ --- коэффициенты нелинейности
(причем в пределе мгновенного нелинейного отклика имеет место отношение
$\beta/\alpha=1/2$). В случае положительных $\alpha$ и $\beta$ (фокусирующая среда)
большой теоретический и практический интерес представляют решения уравнения
(\ref{curl_curl}) в виде т.н. ``световых игл'' --- оптических пучков шириной
порядка длины волны $\lambda_0=2\pi/k_0$, где $k_0=\sqrt{\varepsilon}\omega/c$
\cite{opt_n_0,opt_n_1,opt_n_2,opt_n_3,opt_n_4,circular_polar}. 

Надо сказать, что само существование подобных устойчивых двумерных пространственных
солитонов в рамках уравнения (\ref{curl_curl}) есть результат достаточно нетривиальный,
так как приближение главного порядка для слабонелинейных широких пучков -- система
связанных нелинейных уравнений Шредингера (НУШ) -- не обладает подобными решениями.
Напомним, что если амплитуды $A_\pm(x,y,z)$ левой и правой круговых
поляризаций в выражении
\begin{equation}
{\bf E}\approx \frac{1}{\sqrt{2}}\big[({\bf e}_x+i{\bf e}_y) A_+ 
+ ({\bf e}_x-i{\bf e}_y) A_- \big]e^{ik_0 z} 
\end{equation}
положить медленными функциями и пренебречь малой дивергенцией электрического поля
и его продольной компонентой $E_3$, то получается пара НУШ \cite{BZ1970},
\begin{equation}
-2i k_0 \partial_z A_\pm=\Delta_\perp A_\pm 
+\frac{\alpha k_0^2}{\varepsilon}\Big[|A_\pm|^2 + g|A_\mp|^2\Big]A_\pm,
\label{A_pm_eqs}
\end{equation}
где параметр перекрестной фазовой модуляции $g=1+2\beta/\alpha$. Волновой пакет 
в такой системе либо расплывается за счет дифракции, либо коллапсирует 
за счет нелинейности (см. \cite{Berge1998,ZK2012UFN} и ссылки там). 
Устойчивые солитоны получаются лишь при учете поправок по малому отношению
длины волны к характерной ширине пучка, которые учитывают $\mbox{div}\,{\bf E}\neq 0$
и $E_3 \neq 0$ \cite{MNLSE-1,MNLSE-2,MNLSE-3,beam1,beam2,beam3,beam4,beam5}. 
Для обычных оптических сред с $\varepsilon\sim 1$, где при умеренных интенсивностях 
света всегда выполнено условие $\alpha |{\bf E}|^2\ll \varepsilon$, малыми поправками
можно и ограничиться. Сильные же поля нарушают применимость модели (\ref{curl_curl}),
поскольку включаются неучтенные ею физические процессы \cite{filamentation}.
Поэтому там нет смысла рассматривать предельно узкие, сильно нелинейные солитоны
в рамках уравнения (\ref{curl_curl}).
Но в искусственно созданных композитных материалах могут осуществляться достаточно
малые значения $\varepsilon$ (а также параметр $\gamma=\beta/\alpha$ может 
варьироваться в широких пределах) \cite{composite-1,composite-2,composite-3}. 
В таком случае сильная нелинейность достижима еще при физически умеренной амплитуде
$E=|{\bf E}|$, и есть интерес в получении солитонных решений уравнения (\ref{curl_curl}) 
в непертурбативной области параметров (несколько соответствующих
примеров представлены на рисунке 1).

В недавней работе \cite{circular_polar} были найдены профили циркулярно поляризованных
световых пучков. Такие решения обладают (обобщенной) осевой симметрией --- в полярной
системе координат все три компонетны электрического поля пропорциональны одному и
тому же угловому множителю $\exp(im\varphi)$, так что уравнение (\ref{curl_curl}) 
допускает точную редукцию к системе четырех обыкновенных дифференциальных уравнений 
для функций от поперечной радиальной координаты. При нахождении солитона там пришлось
решать нелинейную краевую задачу с одновременной ``тонкой настройкой'' двух 
действительных параметров. В данной работе рассматриваются линейно поляризованные пучки,
поперечные профили которых не обладают вращательной симметрией (см. пример на рисунке 2), 
и потому приходится использовать совершенно иной метод. 

{\bf Метод решения.}
Мы будем искать такие решения, в которых зависимость электрического поля от 
продольной координаты $z$ сводится к множителю $e^{i\kappa z}$, где $\kappa$
--- константа распространения. Такой множитель будет нами далее подразумеваться. 
Заметим попутно, что чем больше безразмерный (положительный) параметр 
$[\kappa/k_0 -1]$, тем более нелинейным должно быть солитонное решение. 
Поперечные компоненты векторного уравнения тогда выглядят следующим образом:
\begin{eqnarray}
&&-i\kappa(i\kappa E_1-\partial_x E_3)-\partial_y(\partial_y E_1-\partial_x E_2)
\nonumber\\
&&\qquad\qquad\qquad\qquad\qquad-\partial \Pi/\partial E_1^*\equiv f_1=0,
\label{eq_E1}\\
&&-i\kappa(i\kappa E_2-\partial_y E_3)+\partial_x(\partial_y E_1-\partial_x E_2)
\nonumber\\
&&\qquad\qquad\qquad\qquad\qquad-\partial \Pi/\partial E_2^*\equiv f_2=0,
\label{eq_E2}
\end{eqnarray}
где функция $\Pi$ определена выражением
\begin{equation}
\Pi=k_0^2[E^2+(\alpha/\varepsilon) E^4/2+(\beta/\varepsilon)|({\bf E}\cdot{\bf E})|^2/2],
\end{equation}
а ее производные
\begin{equation}
\partial \Pi/\partial {\bf E}^*=k_0^2[(1+(\alpha/\varepsilon) E^2){\bf E}
+(\beta/\varepsilon)({\bf E}\cdot{\bf E}) {\bf E}^*].
\end{equation}

Вместо $z-$компоненты мы воспользуемся следствием уравнения (\ref{curl_curl}), которое
получается применением к нему оператора дивергенции. Таким образом, третье уравнение
нашей системы имеет вид
\begin{equation}
i\kappa (\partial \Pi/\partial E_3^*)+
\mbox{div}_\perp(\partial \Pi/\partial {\bf E}^*_\perp)\equiv p=0.
\label{eq_div}
\end{equation}

Нac здесь интересуют такие решения, в которых функции $E_1(x,y)$ и $E_2(x,y)$ --- 
чисто действительные, тогда как $E_3(x,y)$ --- чисто мнимая. 
Это свойство как раз соответствует линейной поляризации света.

Нетрудно видеть, что уравнения (\ref{eq_E1}) и (\ref{eq_E2}) представляют собой
условия экстремальности функционала
\begin{equation}
F=\int\Big[|i\kappa{\bf E}_\perp-\nabla_\perp E_3 |^2
         +|\partial_y E_1-\partial_x E_2|^2-\Pi\Big] dx dy
\end{equation}
при его варьировании по $\delta {\bf E}^*_\perp$, а уравнение (\ref{eq_div}) есть
равенство нулю вариации функционала $F$ по потенциальной компоненте 
$\delta {\bf E}^*_{\rm pot}=\nabla \delta\Phi^*$.

Из структуры функционала ясно, что решение уравнений является не точкой строгого 
минимума либо максимума $F$, а его седловой точкой, поскольку соленоидальная часть
поперечного поля на малых масштабах дает большой положительный вклад, тогда как
потенциальная часть на малых масштабах дает отрицательный вклад. Казалось бы, 
можно попробовать релаксационную процедуру (со вспомогательной псевдо-эволюционной
переменной $\tau$) вида $d{\bf E}_\perp/d\tau=-{\bf f}_\perp$, 
$d E_3/d\tau=iCp\equiv -f_3$, где $C$ --- достаточно большой положительный
коэффициент. Такая диссипативная ``динамика'' обеспечивает затухание любых
мелкомасштабных возмущений. Но, как показали уже первые попытки реализации, 
в подобной системе псевдо-эволюционных уравнений все равно остается неустойчивая
мода, проявляющаяся на основном масштабе. Чтобы удержать эту единственную 
оставшуюся неустойчивую моду, мы должны вспомнить еще об одном следствии уравнения
(\ref{curl_curl}), а имено об уравнении
\begin{equation}
\mbox{div}\,\mbox{Im}\,[{\bf E}^*\times\mbox{rot} {\bf E}]=0. 
\end{equation}
Оно есть условие равенства нулю дивергенции плотности потока энергии (усредненного 
по периоду колебаний вектора Пойнтинга) в монохроматическом оптическом поле. 
Отсюда следует наличие дополнительного инварианта --- полного потока энергии 
$W$ через поперечное сечение пучка. С точностью до постоянного множителя
\begin{equation}
W=2\mbox{Re}\int\Big[\kappa|{\bf E}_\perp|^2
           +i({\bf E}^*_\perp\cdot\nabla_\perp E_3)\Big]dx dy. 
\end{equation}
Заметим, что учет второго -- непараксиального -- слагаемого под знаком интеграла
становится абсолютно необходим при рассмотрении узких солитонов.
Чтобы обеспечить фиксацию $W$, в псевдо-эволюционной процедуре решались уравнения
\begin{equation}
d{\bf E}/d\tau=-{\bf f}+\mu [\delta W/\delta {\bf E}^*],
\label{relax}
\end{equation}
где вариационная производная
\begin{equation}
[\delta W/\delta {\bf E}^*]=2\kappa  {\bf E}_\perp +i\nabla_\perp E_3
+i{\bf e}_z (\nabla_\perp\cdot {\bf E}_\perp).
\end{equation}
При этом действительная величина $\mu$ самосогласованно изменялась согласно формуле
\begin{equation}
\mu=\int\mbox{Re}({\bf f}\cdot[\delta W/\delta {\bf E}])dx dy
    \Bigg/\int|\delta W/\delta {\bf E}^*|^2 dx dy.
\end{equation}
Выполнением этой процедуры на достаточно большом интервале переменной $\tau$ 
достигается минимум функционала $F$ при локальном дополнительном условии
(\ref{eq_div}) и заданном значении интеграла $W$.

При разумном выборе формы начального состояния (с общей амплитудой $A_0$) 
система (\ref{relax}) остается на ``гиперповерхности'' $W=W_0\propto A_0^2$ 
и приближается вдоль нее к устойчивой предельной точке, которая может оказаться
близкой к интересующему нас солитонному решению. 
Асимптотическая величина $\mu_\infty$ может оказаться как положительной, 
так и отрицательной --- в зависимости от значения $W_0$, но нас не устраивает
ни один из этих вариантов. А нужен нам как раз тот специальный случай,
когда  $\mu_\infty(W_0)=0$  (так что в пределе и ${\bf f}(x,y)\to 0$, что есть 
наша главная цель). Остается только подобрать оптимальное $A_0$, используя 
``метод стрельбы''. В результате тонкой настройки всего одного параметра мы получаем 
искомое решение для профиля светового пучка при заданном значении $\kappa> k_0$.

Эволюционный шаг по $\tau$ выполнялся в соответствии со схемой Рунге-Кутта
4-го порядка аппроксимации с применением псевдо-спектрального метода по поперечным
координатам, на сетке в $320\times 320$ точек. Насчет точности и эффективности
предложенной процедуры надо сказать, что за разумное время порядка часа на
персональном компьютере удается добиться достаточно малой среднеквадратичной
ошибки $|{\bf f}|\sim 10^{-5}$.

\begin{figure}
\begin{center} 
\epsfig{file=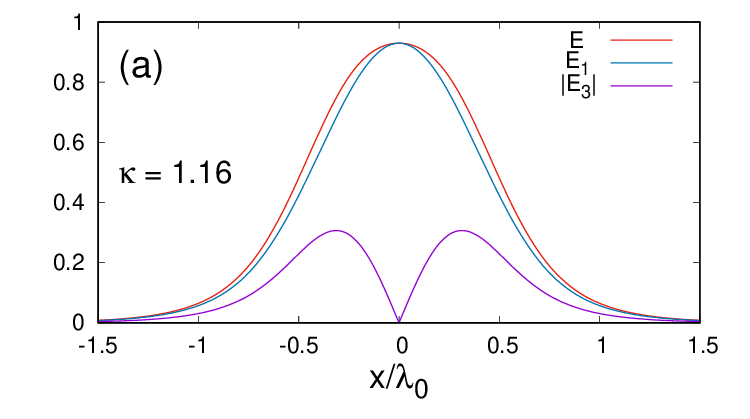, width=84mm}\\
\epsfig{file=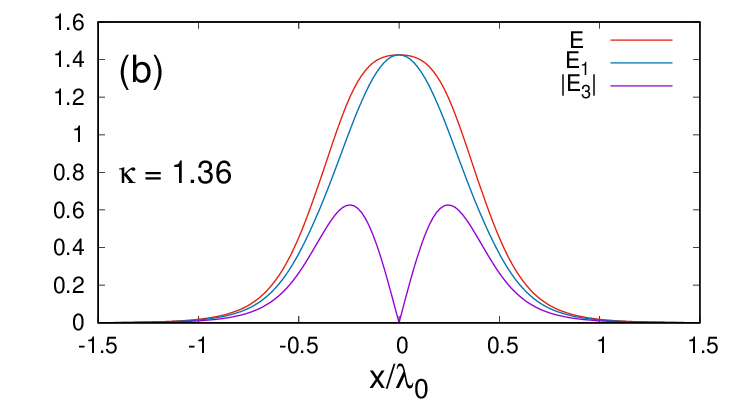, width=84mm}\\
\epsfig{file=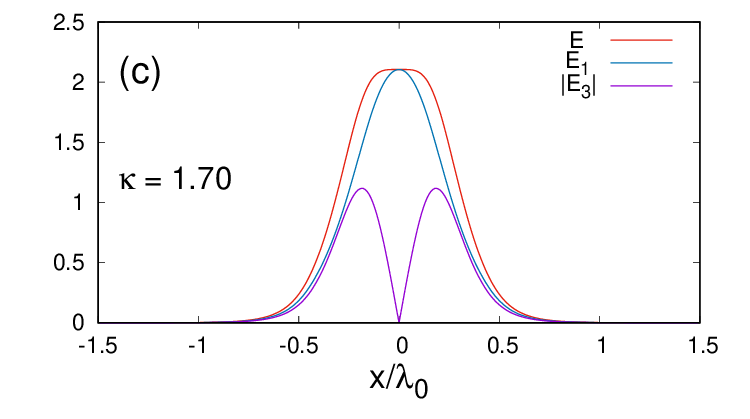, width=84mm}
\end{center}
\caption{Примеры профилей линейно поляризованных керровских солитонов
большой амплитуды при $y=0$ для трех значений рараметра $\kappa$. 
Компонента $E_2$ на этой линии равна нулю (см. подробности в тексте).
}
\label{E_E1_E3_x} 
\end{figure}

\begin{figure}
\begin{center} 
\epsfig{file=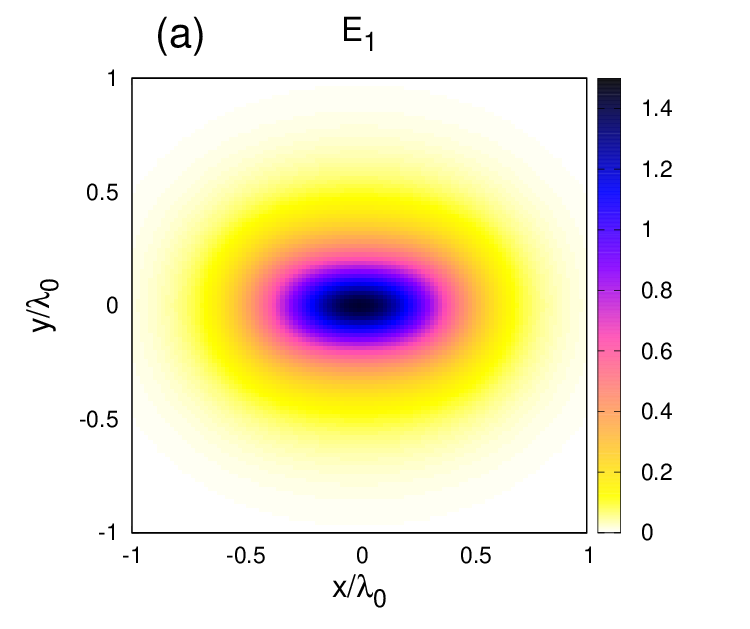, width=72mm}\\
\epsfig{file=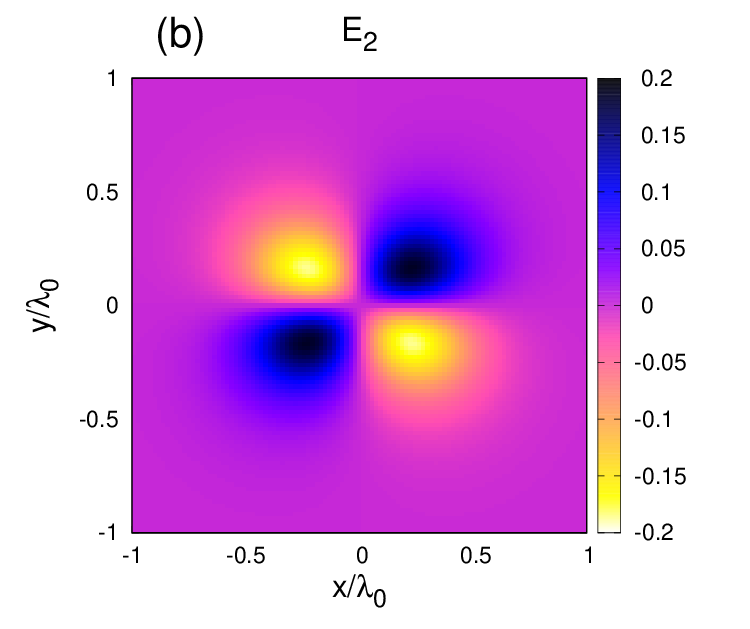, width=72mm}\\
\epsfig{file=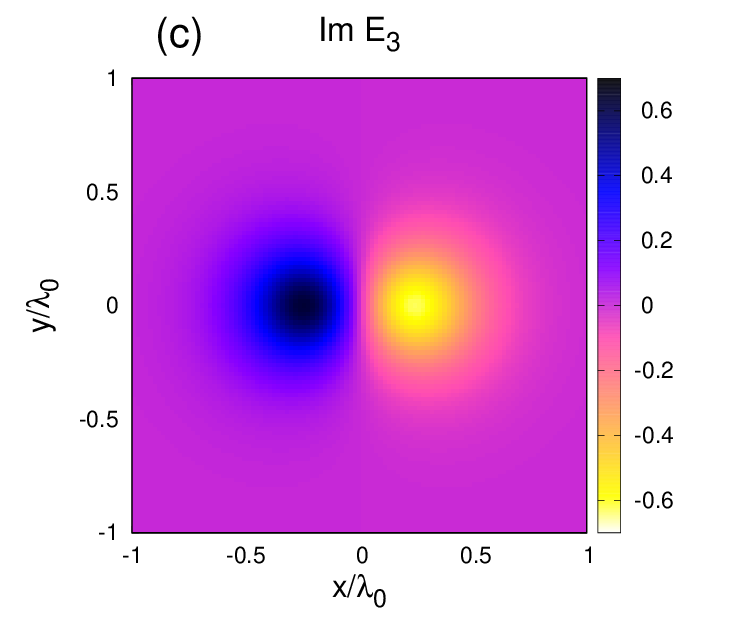, width=72mm}
\end{center}
\caption{Компоненты электрического поля как функции поперечных координат при 
$\kappa=1.36$. Показана не вся вычислительная область, а только ее центральная часть. 
}
\label{E123_xy} 
\end{figure}

{\bf Численные примеры.}
Примеры полученных таким способом сильно нелинейных солитонов приведены
на рисунках 1-2 в терминах обезразмеренных величин, что формально соответствует
параметрам $k_0=1$, $\alpha/\varepsilon=1$, $\beta/\alpha=0.5$. 

Начальное состояние при $\tau=0$ выбиралось в виде суммы двух изотропных 
гауссианов для $E_1$, а остальные две компоненты были равны нулю. 
В процессе псевдо-эволюции ``вырастали'' нетривиальные $E_2$ и $E_3$, но они
оставались заметно меньшими по сравнению с $E_1$. Это приводило в конце к 
поляризованному в среднем вдоль оси $x$ световому пучку. 
На рисунке 1 представлены зависимости компонент электрического поля на линии симметрии
$y=0$ для трех различных значений константы распространения. Видно, что с увеличением
$\kappa$ отношение $(\mbox{max}\,|E_3|/\mbox{max}\,|E_1|)$ существенно увеличивается.
Ширина солитона при этом становится не просто сравнимой, а даже заметно меньшей,
чем длина волны. Чтобы не потерять точность аппроксимации, в случае 1(c) размеры
вычислительной области были уменьшены до $3\lambda_0\times 3\lambda_0$ по сравнению с
$6\lambda_0\times 6\lambda_0$ в случаях 1(a) и 1(b).

На рисунке 2 показано распределение всех трех компонент поля в поперечной плоскости
при $\kappa=1.36$. Видно, что компонента $E_1$ не имеет нулевых линий, компонента
$E_3$ (чисто мнимая) обращается в ноль на прямой $x=0$, а компонета $E_2$ обращается
в ноль на прямых $x=0$ и $y=0$. Еще отметим бросающуюся в глаза на рисунке 2(a) 
анизотропию функции $E_1(x,y)$ --- сечение пучка несколько сплюснуто поперек
доминирующего направления поляризации. Ширина солитона вдоль направления оси $y$
может составлять менее половины длины волны.

{\bf Заключение.} 
Таким образом, в данной работе выяснена детальная структура линейно поляризованных 
оптических пучков большой амплитуды в керровской среде. Подобные решения невозможно
получить пертурбативными методами. Успешное применение по сути вариационной
численной процедуры говорит об устойчивости этих объектов по отношению к трехмерным 
(чисто пространственным) возмущениям. Наличие еще хотя бы одной неустойчивой моды
неизбежно проявилось бы в псевдо-эволюционной процедуре и не позволило бы системе
попасть в предельную точку.

В будущем можно попробовать применить этот метод также и к моделям с насыщающейся
нелинейностью, в которых коэффициенты $\alpha$ и $\beta$ уменьшаются с увеличением
интенсивности $E^2$ \cite{saturat-1,saturat-2,saturat-3}.

{\bf Финансирование работы.}
Работа выполнена в рамках госзадания по теме FFWR-2024-0013.

{\bf Конфликт интересов.}
Автор данной работы заявляет, что у него нет конфликта интересов.


\begin{thebibliography}{99}

\bibitem{LL8} Л. Д. Ландау и Е. М. Лифшиц, {\it Электродинамика сплошных сред},
Наука, Москва (1982).

\bibitem{opt_n_0} D. Pohl, Opt. Commun. {\bf 2}, 305 (1970).

\bibitem{opt_n_1} В. Е. Семенов, Н. Н. Розанов, Н. В. Высотина,
ЖЭТФ {\bf 116}, 458 (1999).

\bibitem{opt_n_2} N. N. Rosanov, V. E. Semenov and N. V. Vyssotina,
J. Opt. B: Quant. Semiclass. Opt. {\bf 3}, 96 (2001). 

\bibitem{opt_n_3}
N. N. Rosanov, V. B. Semenov, N. A. Solov'eva, N. V. Vyssotina,
Proc. SPIE {\bf 4751}, 1 (2002).

\bibitem{opt_n_4} В. П. Рубан, Письма в ЖЭТФ {\bf 120}, 745 (2024).

\bibitem{circular_polar} В. П. Рубан, Письма в ЖЭТФ {\bf 121}, 375 (2025).

\bibitem{BZ1970} А. Л. Берхоер, В. Е. Захаров, ЖЭТФ {\bf 58}, 903 (1970).

\bibitem{Berge1998} L. Berge, Phys. Rep. {\bf 303}, 259 (1998).

\bibitem{ZK2012UFN} В. Е. Захаров, Е. А. Кузнецов, 
Усп. Физ. Наук {\bf 182}, 569 (2012).

\bibitem{MNLSE-1} S. Chi and Q. Guo, Opt. Lett. {\bf 20}, 1598 (1995).

\bibitem{MNLSE-2}  B. A. Malomed, K. Marinov, D. I. Pushkarov, and A. Shivarova,
Phys. Rev. A {\bf 64}, 023814 (2001).

\bibitem{MNLSE-3} G. Fibich and B. Ilan, Phys. Rev. E {\bf 67}, 036622 (2003).

\bibitem{beam1} Н. Н. Розанов, Н. В. Высотина, А. Г. Владимиров,
ЖЭТФ {\bf 118}, 1307 (2000).

\bibitem{beam2} N. N. Rosanov, Proc. SPIE {\bf 4403}, 200 (2001).

\bibitem{beam3} Н. Н. Розанов, Опт. Спектроск. {\bf 94}, 1013 (2003).

\bibitem{beam4} N. V. Vysotina, N. N. Rozanov, V. E. Semenov, V. A. Smirnov,
S. V. Fedorov, and D. N. Christodoulides, Optics and Spectroscopy {\bf 98}, 895 (2005).

\bibitem{beam5} H. Wang and W. She, Opt. Expr. {\bf 13}, 6931 (2005). 

\bibitem{filamentation} A. Couairon, A. Mysyrowicz, 
Phys. Rep. {\bf 441}, 47 (2007).

\bibitem{composite-1} J. E. Sipe and R. W. Boyd, Phys. Rev. A {\bf 46}, 1614 (1992).

\bibitem{composite-2} A. Ciattoni, C. Rizza, and E. Palange,
Phys. Rev. A {\bf 81}, 043839 (2010).

\bibitem{composite-3} C. Rizza, A. Ciattoni, and E. Palange,
Phys. Rev. A {\bf 83}, 053805 (2011).

\bibitem{saturat-1} N. V. Vysotina, N. N. Rozanov, V. E. Semenov, V. A. Smirnov,
and S. V. Fedorov, Optics and Spectroscopy {\bf 98}, 447 (2005).

\bibitem{saturat-2} F. Bouchard, H. Larocque, A. M. Yao, 
C. Travis, I. De Leon, A. Rubano, E. Karimi, G.-L. Oppo, and R. W. Boyd,
Phys. Rev. Lett. {\bf 117}, 233903 (2016).

\bibitem{saturat-3} C. J. Gibson, P. Bevington, G.-L. Oppo, and A. M. Yao,
Phys. Rev. A {\bf 97}, 033832 (2018).

\end{thebibliography}
\end{document}